\documentclass[%
reprint,
superscriptaddress,
amsmath,amssymb,
aps,
floatfix,
]{revtex4-1}

\usepackage[english]{babel}
\usepackage[dvips]{graphicx}
\usepackage{color}
\usepackage{float}
\usepackage{latexsym}
\usepackage{amsmath}
\usepackage{amsthm}
\usepackage{amsfonts}
\usepackage{amssymb}
\usepackage{romannum}
\usepackage[T1]{fontenc}
\usepackage{braket}
\usepackage{dsfont}
\usepackage{xcolor}
\usepackage{bm}

\usepackage{soul}

\begin{document}
	
\pagenumbering{arabic}
	
\title{Geometrical Rabi oscillations and Landau-Zener transitions in non-Abelian systems}
\author{H. Weisbrich}
\affiliation{Fachbereich Physik, Universit{\"a}t Konstanz, D-78457 Konstanz, Germany}
\author{G. Rastelli}
\affiliation{Fachbereich Physik, Universit{\"a}t Konstanz, D-78457 Konstanz, Germany}
\affiliation{INO-CNR BEC Center and Dipartimento di Fisica, Universit{\`a} di Trento, I-38123 Povo, Italy}
\author{W. Belzig}
\email{Corresponding author: wolfgang.belzig@uni-konstanz.de}
\affiliation{Fachbereich Physik, Universit{\"a}t Konstanz, D-78457 Konstanz, Germany}

\begin{abstract}
Topological phases of matter became a new standard to classify quantum systems in many cases, yet key quantities like the quantum geometric tensor providing local information about topological properties are still experimentally hard to access.
In non-Abelian systems this accessibility to geometric properties can be even more restrictive due to the degeneracy of the states.
We propose universal protocols to determine quantum geometric properties in non-Abelian systems. First, we show that for a weak resonant driving of the local parameters the coherent Rabi oscillations are related to the quantum geometric tensor. Second, we derive that in a Landau-Zener like transition the final probability of an avoided energy crossing is proportional to elements of the non-Abelian quantum geometric tensor. Our schemes suggest a way to prepare eigenstates of the quantum metric, a task that is difficult otherwise in a degenerate subspace.
\end{abstract}

\date{\today}

\maketitle
\textit{Introduction.}--The geometry of quantum states is crucial in many branches of physics. It scopes the field of the Aharonov-Bohm effect \cite{aharonov1959significance,wu1975concept}, the Berry phase \cite{berry1984quantal,wilczek1984app} and more recently the concept of topological phases such as topological insulators \cite{Hasan:2010ku}, topological semi-metals \cite{armitage2018weyl} and topological superconductors \cite{Sato:2017go}.
The key quantity related to these phenomena is the quantum geometric tensor (QGT). On one hand, its real part yields the quantum metric that quantifies the distance between different quantum states \cite{anandan1990geometry}. This general property can be connected to a wide spectrum of physical phenomena. For instance it is essential for understanding superfluidity in flat bands \cite{julku2016geometric}, orbital magnetic susceptibility \cite{gao2014field,piechon2016geometrical}, the anomalous Hall effect \cite{gao2014field,bleu2018effective,gianfrate2020measurement}, and quantum phase transitions \cite{venuti2007quantum,zanardi2007information}. Besides it was used to determine the topological invariant of a Tensor monopole \cite{tan2021experimental,palumbo2018revealing}, and it also defines the Euler number, a topological invariant characterizing non-trivial topology in gapped fermionic systems \cite{ma2013euler}.
On the other hand, the imaginary part of the QGT yields the Berry curvature. This curvature is related to the geometric phase accumulated along a cyclic path yielding the Berry phase \cite{berry1984quantal} or similar in an electromagnetic gauge potential the Aharonov-Bohm effect \cite{aharonov1959significance,wu1975concept}. Moreover the integration of the Berry curvature over a closed two-dimensional manifold defines the first Chern number, the topological invariant for a wide spectrum of phenomena, such as the quantum Hall effect \cite{TKKN,klitzing1980new} or conducting edge states in topological insulators \cite{Hasan:2010ku}.

The accessibility to the QGT is thus crucial to analyze many recently studied phenomena in physics. There are several proposal to measure the Abelian geometric properties, for instance the quantum metric can be extracted by quantum quenches \cite{kolodrubetz2013classifying}, by analyzing the current noise \cite{neupert2013measuring}, or in photonic system \cite{bleu2018measuring}. Another approach is via periodic driving to extract the Abelian QGT \cite{klees2020microwave,ozawa2018extracting}. The latter also realized experimentally in a superconducting qubit \cite{tan2019experimental} and a set of coupled qubits in diamond \cite{yu2020experimental}.
Fundamental physical phenomena related to geometry, however, are not restricted to the Abelian case, for instance non-Abelian Majorana zero modes in topological superconductors promise protected quantum computation \cite{sarma2015majorana,deng2016majorana,aasen2016milestones,karzig2017scalable}. Also in other non-Abelian cases, as in the 4D quantum Hall \cite{zhang2001four,price2015four,kraus2013four,lohse2018exploring,zilberberg2018photonic}, in other systems with a non-trivial second Chern number \cite{sugawa2018second,lu2018topological,weisbrich2021second}, or in holonomic quantum computation \cite{pachos1999non,zanardi1999holonomic}, non-Abelian geometry plays an extraordinary role and it is essential to understand underlying mechanism. However extracting the full QGT in the non-Abelian case stays elusive.

In this Letter, we report a universal approach to extract the QGT in non-Abelian systems using coherent Rabi oscillation by driving the system periodically. 
We show that for a system depending on a set of parameters defining the geometry of the problem this goal can be achieved by either modulating a single parameter or by a modulation of two parameters.  
Our proposal does not require any adiabatic condition and can be used to prepare eigenstates of the QGT in the degenerate subspace.
We also show that our approach shows that the rates in a Landau-Zener-like transition are directly determined by the elements of the QGT.

\textit{Single parameter modulation.}--
We assume a two-band  model  
$H_0(\bm{\lambda})=\sum_{\sigma=\pm} E_{\sigma}(\bm{\lambda})\sum_{\nu=1}^{N} \ket{\psi_{\nu}^{\sigma}(\bm{\lambda})}\bra{\psi_{\nu}^{\sigma}(\bm{\lambda})}$, 
with $\ket{\psi_\nu^\pm}$ the states of the degenerate energy levels $E_{\pm}$ and degeneracy dimension $N$. Furthermore the Hamiltonian does depend on $M$ external dimensionless parameter $\bm{\lambda}=(\lambda_1,...,\lambda_M)$. By a modulation of the external parameter $\lambda_j$ 
\begin{align}
\lambda_j\rightarrow \lambda_j+\frac{2A}{\hbar\omega}\cos(\omega t)
\end{align}
with a small amplitude $A/\hbar\omega\ll1$ the Hamiltonian can be expanded to first order in the driving to
\begin{align}
H_1=H_0(\bm{\lambda})+\frac{2A}{\hbar\omega}\cos(\omega t)\partial_{\lambda_j}H_0(\bm{\lambda})\, .
\end{align}
In the rotating frame of the drive and using the rotating wave approximation (RWA) this results in 
\begin{align}
H_1^\textrm{RWA}&=\sum_{\nu=1}^{N} \sum_{\sigma=\pm} 
\left(E_\sigma -\sigma\hbar\omega/2\right)\ket{{\psi}_\nu^{\sigma}}\bra{{\psi}_\nu^{\sigma}}\nonumber\\ &-A\sum_{\nu,\mu=1}^{N}\left(\ket{{\psi}_\nu^-}\braket{\partial_j{\psi}_\nu^-|{\psi}_\mu^+}\bra{{\psi}_\mu^+}+h.c.\right) \, . \label{eq:3}
\end{align}
as derived in the Supplemental Material \cite{supl_mat}.
Solving the Schr{\"o}dinger equation 
$i\hbar\partial_t\ket{\Psi}=H_{1}^{RWA}\ket{\Psi}$ 
in the basis 
$\ket{\Psi}=\sum_{\nu=1}^{N} \sum_{\sigma= \pm} c_{\nu}^{\sigma}(t)e^{-i(E_{\sigma}/\hbar-\sigma\omega/2)t}\ket{\psi_{\nu}^{\sigma}}$ we find that \cite{supl_mat}
\begin{align}
\label{eq:dyn}
\ddot{\bm{c}}^{\pm}  =-\frac{A^2}{\hbar^2}Q_{jj}^\pm\bm{c}^\pm \pm i\delta\omega \, \dot{\bm{c}}^\pm 
\end{align}
with $\bm{c}^\pm=(c_{1}^\pm,...,c_{N}^\pm)$, the detuning $\delta\omega =(E_{+}-E_{-})/\hbar-\omega$ with $|\delta\omega| \ll \omega$, 
and the non-Abelian QGT 
\begin{equation}
[Q_{jk}^\pm]_{\nu\mu}=\bra{\partial_{\lambda_{j}}\psi_{\nu}^\pm}\left(\mathds{1}-\sum_{\alpha=1}^{N}\ket{\psi_{\alpha}^\pm}\bra{\psi_\alpha^\pm}\right)\ket{\partial_{\lambda_{k}}\psi_{\mu}^\pm} \, .
\end{equation}
Notice that in Eq.~(\ref{eq:dyn}) we used the form of a second-order differential equation in time in which $\bm{c}^{+}$ and $\bm{c}^{-}$ are decoupled.
\begin{figure}[t]
	\includegraphics[width=0.45\textwidth]{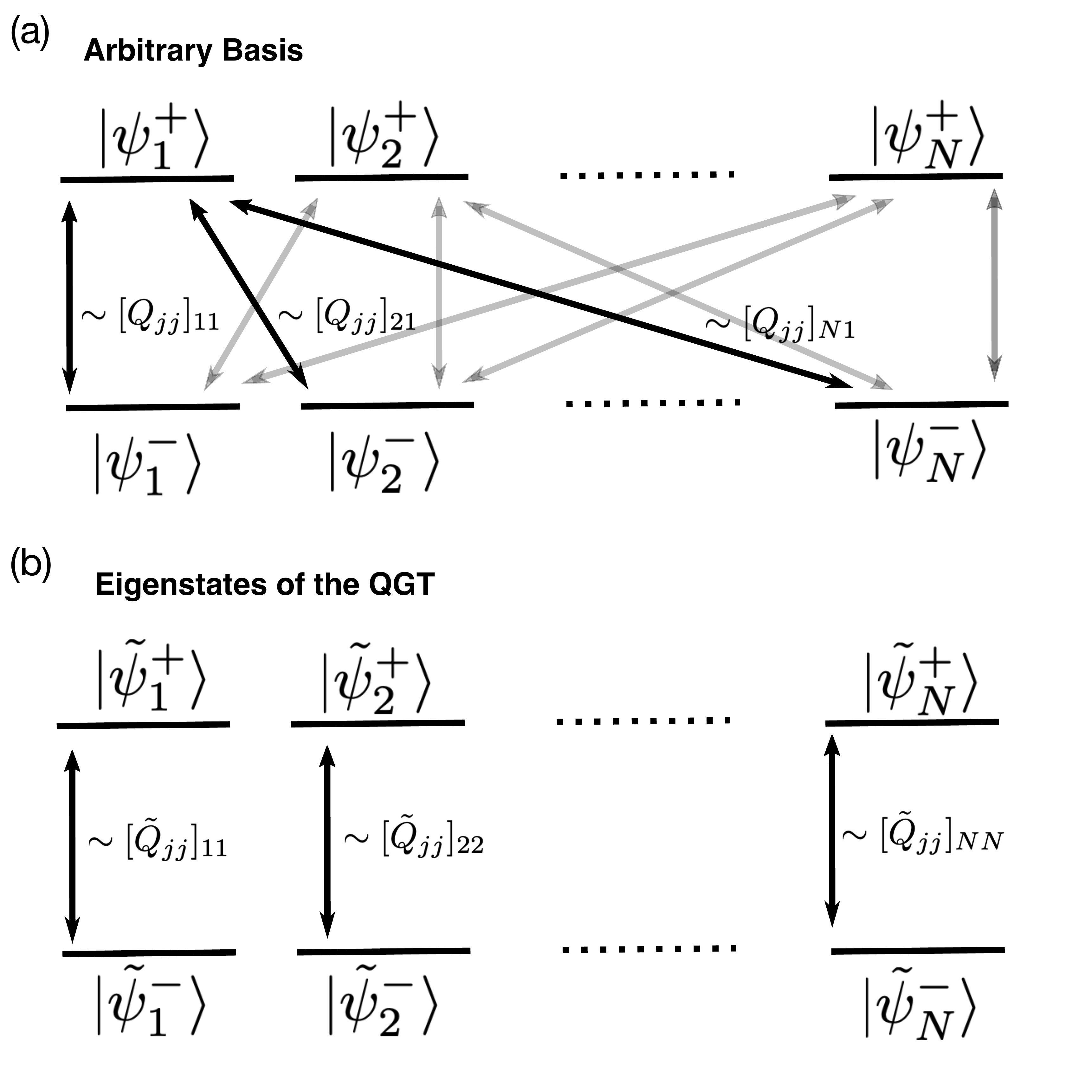}
	\caption{Schematic of the interaction between the states of the two energy levels by a modulation of a single external parameter in a non-Abelian system in an arbitrary basis (a) and in the basis of the eigenstates of the QGT (b). }\label{Fig1}
\end{figure}
In an arbitrary basis the QGT is not diagonal, thus coupling different states of the degenerate energy levels, see Fig.~\ref{Fig1}(a). 
However the QGT can be also diagonalized in the eigenbasis and we can write 
$\ket{\Psi}=\sum_{\nu,\sigma=\pm}\tilde{c}_\nu^\sigma e^{-i(E_{\sigma}/\hbar-\sigma\omega/2)t}\ket{\tilde{\psi}_{\nu}^\sigma}$ such that the diff. equation simplifies to
 \begin{align}
\ddot{\tilde{c}}_{\nu}^\pm=-\frac{A^2}{\hbar^2}[\tilde{Q}_{jj}]_{\nu\nu}\tilde{c}_\nu^\pm \pm i\delta\omega \, \mathds{1}\dot{\tilde{c}}_\nu^\pm\label{eq:4} \, ,
\end{align}
with the diagonalized QGT $\tilde{Q}_{jj}:=\tilde{U}^\pm Q_{jj}^\pm (\tilde{U}^\pm)^\dagger$, and $\tilde{U}^\pm(\ket{\psi^\pm_1},...,\ket{\psi^\pm_N})=(\ket{\tilde{\psi}^\pm_1},...,\ket{\tilde{\psi}^\pm_N})$. 
Hence each pair of eigenstates $\ket{\tilde{\psi}_\nu^\pm}$ of the QGT oscillates with a Rabi frequency proportional to its eigenvalue of the QGT $[\tilde{Q}_{jj}]_{\nu\nu}$, as depicted in Fig.~\ref{Fig1}(b).
This essentially resembles the Morris-Shore transformation \cite{morris1983reduction,shore2014two}, which transforms two degenerate interacting bands into independent interacting two-state systems by a basis transformation diagonalizing the perturbation/interaction. 
In the resonant case $\delta\omega=0$ the Rabi frequency between a pair of eigenstates $\ket{\tilde{\psi}_\nu^\pm}$ can be directly extracted from Eq.~\ref{eq:4} as
\begin{align}
\Omega^{s}_\nu=\frac{A}{\hbar}[Q_{jj}]_{\nu\nu}^{1/2}
\end{align}
solving Eq.~\ref{eq:4} and the Schr{\"o}dinger equation with   
$\tilde{c}_\nu^+(t)=-i\sin(\Omega^s_\nu t)$ and $\tilde{c}_\nu^-(t)=\cos(\Omega^s_\nu t)$ if we start in the lower energy level. 
For an arbitrary initial state the systems oscillations between the lower and upper band is a superposition of several two-states Rabi oscillations, whereas each Rabi frequency is proportional to a different eigenvalues of the QGT $\tilde{Q}_{jj}$.
Hence the oscillations between the bands heavily depend on the initial condition of the system. 
Only if the system is initialized in one of the eigenstates of the QGT a simple two state Rabi oscillation occurs, whereas for an arbitrary initial state not being an eigenstate of the QGT the system will oscillate in a more complex behavior due to the overlapping oscillations with different frequencies.
Then, the straightforward option is to measure the spectrum of the Rabi frequencies to extract the QGT in the diagonal basis.

\begin{figure}[t!]
	\includegraphics[width=0.49\textwidth]{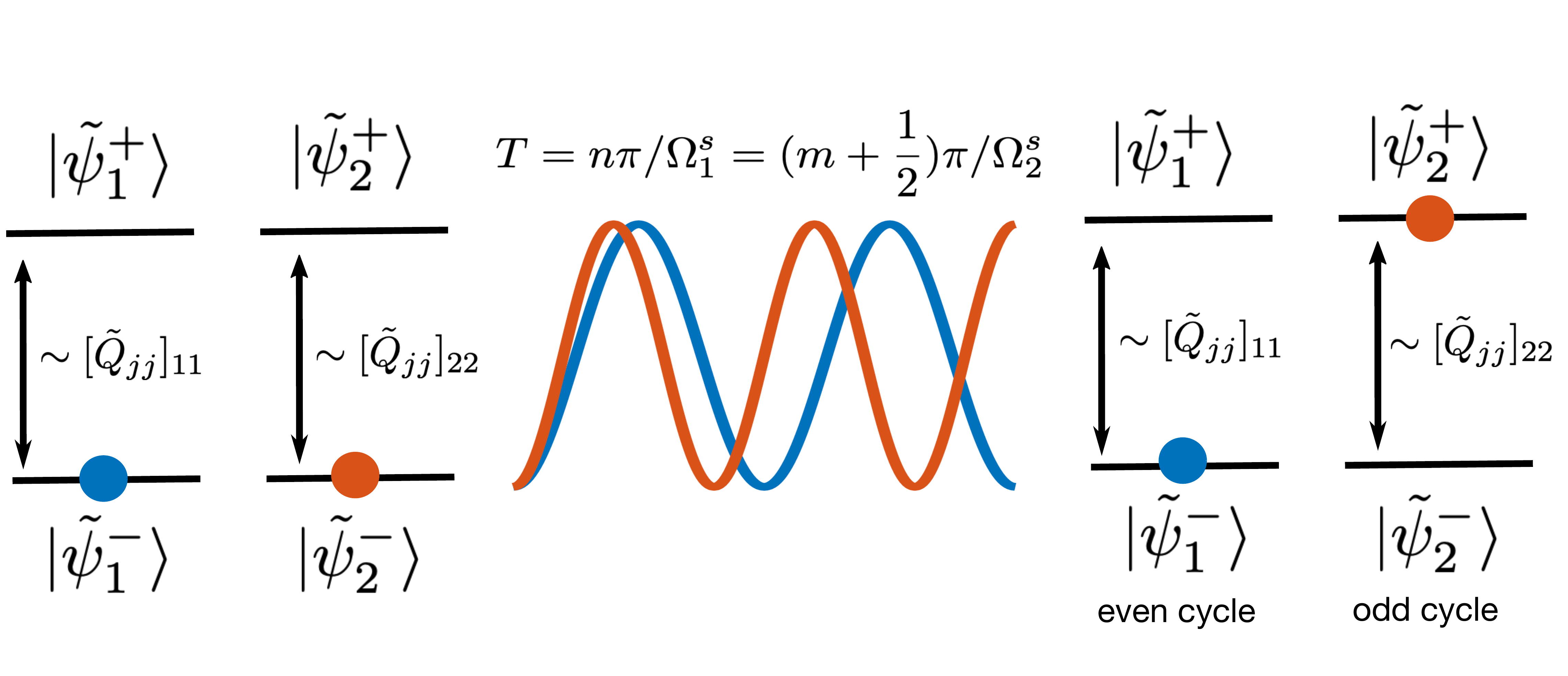}
	\caption{Protocol for preparing into the eigenstates of the QGT starting from an arbitrary state in the ground state. The states evolve for the time $T= n(\pi/\Omega_{1}^{s})= (m+1/2)(\pi/\Omega_{2}^{s})$ yielding an even Rabi cycle for the first pair and an odd Rabi cycle for the second pair due to the different Rabi frequencies $\Omega_1^s\neq \Omega_2^s$ for each pair.}
	\label{Fig2}
\end{figure}

Furthermore one can also exploit the multi-harmonic dynamics to prepare the system in an eigenstate of the QGT.
For instance, for twofold degenerate energy levels the two pairs of eigenstates oscillate with different frequency $\Omega_{1}^s$ and $\Omega_2^s$.
For an arbitrary initial ground state the system will start to oscillate between the two pairs of eigenstates for as long as the interaction is active. 
Hence one chooses a pulse duration $T$ for the drive, such that $T= n(\pi/\Omega_{1}^{s}) = (m+1/2)(\pi/\Omega_{2}^{s})$ with $n,m$ being integers.
On the one hand, there will be an even cycle for the oscillation in the first pair of eigenstates, such that after the duration $T$ the state of this two state system will be again in the lower eigenstate.
On the other hand there is an odd cycle for the second pair of eigenstates resulting in the upper eigenstate after the time $T$. 
With this superposition, by a measurement of the energy after the pulse the state will be an eigenstate of the QGT 
either $\ket{\tilde{\psi}_1^-}$ or $\ket{\tilde{\psi}_2^+}$ depending on the outcome of the measurement, see schematically in Fig.\ref{Fig2}.

However the condition $T= n \pi/\Omega_{1}^{s} = (m+1/2)\pi/\Omega_{2}^{s}$ can be only satisfied exactly for commensurate frequencies. In general, this can be only satisfied approximately, thus having a reduced fidelity of the final state for incommensurate frequencies. In this case, a higher accuracy of the final state can be achieved by a longer pulse duration. For example in case of $\Omega_{2}^{s}/\Omega_{1}^{s}=\pi$ a fidelity of $F=|\braket{\Psi_f|\Psi_t}|^2\approx 97.3\%$ can be achieved by choosing $T=3\pi/\Omega_{1}^s$ with $\Psi_f$ the final state after the pulse and $\Psi_t=(\ket{\tilde{\psi}_1^-}+\ket{\tilde{\psi}_2^+})/\sqrt{2}$ the target state (assuming that the initial state was $\ket{\Psi_i}=(\ket{\tilde{\psi}_1^-}+\ket{\tilde{\psi}_2^-})/\sqrt{2}$), whereas a fidelity of $F\approx 99.2\%$ can be expected at the cost of a longer pulse duration $T=25\pi/\Omega_{1}^s$.

This protocol can be also expanded for $N$-fold degenerate bands. 
Here one can repeat the procedure of a pulse with duration $T$ followed by a measurement of the energy. 
By choosing two sets of eigenstates, e.g. for fourfold degenerate bands the first two eigenstate pairs in the first set and the other two in the second set, one can restrict the state after each measurement to one set/subspace. 
The pulse duration in each step has to fulfill $T=n_{1}(\pi/\Omega_1^{s})\approx n_{2}(\pi/\Omega_2^{s})\approx...\approx n_{N_\nu}(\pi/\Omega_{N_1}^{s})\approx(m_{1}+1/2)(\pi/\Omega_{N_1+1}^s)\approx...\approx(m_{N_2}+1/2)(\pi/\Omega_{N_1+N_2}^s)$ with $N_1$ the number of eigenstates in the first set and $N_2$ the amount of eigenstates in the second set. For increasing dimensions $N$ of the energy levels also the time duration of the pulse would increase drastically, 
but in principle the pulse can be applied similar as in the $N=2$ case.
Hence after a measurement the final state is either a state in the first set (when $E_-$ is the outcome) or otherwise from the second set. 
By repeating this procedure one finally ends up with two set only containing a single pair of eigenstates, where the procedure as discussed in Fig.\ref{Fig2} can be applied to prepare the system in an eigenstate of the QGT.

\textit{Two parameter modulation.}--So far we only discussed the case of a single parameter modulation yielding Rabi oscillations proportional to $Q_{jj}^\pm$. 
For a two parameter drive one is also able to extract $Q_{jk}^\pm$ with the corresponding symmetric quantum metric tensor $g_{jk}=(Q_{jk}+Q_{jk}^\dagger)/2$, 
and the anti-symmetric Berry curvature $F_{jk}=i(Q_{jk}-Q_{jk}^\dagger)$ \cite{ma2010abelian}. 
In  case of a modulation of two parameter the Hamiltonian to first order in the driving reads
\begin{equation}
H_{2}\!
=\!
H_0(\bm{\lambda}) \! + \! 
\frac{2A}{\hbar\omega} \! \big[ \!\!\cos(\omega t)\partial_{\lambda_j}\!H_0(\bm{\lambda})+\cos(\omega t+\phi)\partial_{\lambda_k}\!H_0(\bm{\lambda}) \! \big]
\end{equation}
where we added a phase $\phi$ between the two modulations. Again solving the Schr{\"o}dinger Eq. of the RWA of the Hamiltonian $H_{2}^{RWA}$ in the same basis as for a single modulation yields \cite{supl_mat}
\begin{align}
\ddot{\bm{c}}^\pm \!\!=\!\! 
-\frac{A^2}{\hbar^2}\left(Q_{jj}^{\pm}+Q_{kk}^{\pm}+e^{\pm i\phi}Q_{jk}^\pm+e^{\mp i\phi}Q_{kj}^\pm\right)\bm{c}^\pm 
\!\pm\! 
i\delta\omega \dot{\bm{c}}^\pm
.
\label{eq:7}
\end{align}
For a circular modulation $\phi=\pi/2$ one can simplify $e^{\pm i\phi}Q_{jk}^\pm+e^{\mp i\phi}Q_{kj}^\pm=\pm F_{jk}^{\pm}$ to the non-Abelian Berry curvature. 
Again the system can be diagonalized in the perturbation, however this time it is not related to the eigenbasis of the QGT, as several different contributions appear, each with a different eigenbasis in general. 
In the diagonal basis of $\bar{Q}_{jj}^\pm+\bar{Q}_{kk}^\pm\pm\bar{F}_{jk}^\pm:=\bar{U}^\pm(Q_{jj}^\pm+Q_{kk}^\pm\pm F_{jk}^\pm)(\bar{U}^\pm)^\dagger$ (note that only the sum is diagonal with the same eigenvalues for the cases of $\pm$ but not necessarily each individual contribution)
 with $\bar{U}^\pm(\ket{\psi^\pm_1},...,\ket{\psi^\pm_N})=(\ket{\bar{\psi}^\pm_1},...,\ket{\bar{\psi}^\pm_N})$  
 the equation reads
\begin{align}
\ddot{{\bar{c}}}^\pm_\nu=-\frac{A^2}{\hbar^2}\left[\bar{Q}_{jj}^{\pm}+\bar{Q}_{kk}^{\pm}\pm\bar{F}_{jk}^\pm\right]_{\nu\nu}{\bar{c}}^\pm_\nu \pm i\delta\omega\, \dot{\bar{c}}^\pm_\nu\, .
\end{align}
Thus in the resonant case $\delta\omega=0$ the Rabi-frequencies are related to the eigenvalues of $(Q_{jj}^{\pm}+Q_{kk}^{\pm}\pm F_{jk}^\pm)$, where each pair $\ket{\bar{\psi}_\nu^\pm}$ oscillates with its respective frequency
\begin{align}
\Omega_{\nu}^{\pi/2}=\frac{A}{\hbar}\left[\bar{Q}_{jj}^{\pm}+\bar{Q}_{kk}^{\pm}\pm\bar{F}_{jk}^\pm\right]_{\nu\nu}^{1/2}
\end{align}
similar as for the single parameter modulation.

Besides the same protocol as discussed in Fig.~\ref{Fig2} for the state preparation can be applied here. 
However, to get access to the Berry curvature and not only the eigenvalues of the sum $\bar{Q}_{jj}^{\pm}+\bar{Q}_{kk}^{\pm}\pm\bar{F}_{jk}^\pm$, one needs to determine $\bar{Q}_{jj}^\pm$ (and similar $\bar{Q}_{kk}^\pm$) in the new eigenbasis of the two-parameter modulation. 
If these are known the condition $\frac{A^2}{\hbar^2}\left[\bar{Q}_{jj}^{\pm}+\bar{Q}_{kk}^{\pm}\pm\bar{F}_{jk}^\pm\right]_{\nu\mu}=\delta_{\nu\mu}(\Omega^{\pi/2}_{\nu})^2$ can be solved for the Berry curvature $\bar{F}_{jk}^\pm$. 
Since $\tilde{Q}_{jj}^\pm$ is known from the single modulation in the diagonal basis, it is sufficient to determine transformation $\tilde{U}^\pm(\bar{U}^\pm)^\dagger$ with $\bar{Q}_{jj}=\bar{U}^\pm(\tilde{U}^\pm)^\dagger\tilde{Q}_{jj}^\pm\tilde{U}^\pm(\bar{U}^\pm)^\dagger$.

To illustrate the idea,
we  discuss how one can determine this transformation for the simplest case of $N=2$. 
First one has to prepare the system in an eigenstate of the QGT (as discussed in Fig.\ref{Fig2})
let's say $\ket{\tilde{\psi}_\nu^{-}}=a_{\nu 1}^{-} \ket{\bar{\psi}_1^{-}}+a_{\nu 2}^{-} \ket{\bar{\psi}_2^{-}}$ 
with the complex coefficients $a_{\nu\mu}^{-}$ defining the transformation
 $[\tilde{U}^{-}(\bar{U}^{-})^\dagger]_{\nu\mu}=a_{\nu\mu}^{-}$. 
 We then apply a pulse with the two parameter modulation for the time $T=n(\pi/\Omega_1^{\pi/2})\approx (m+1/2)(\pi/\Omega_2^{\pi/2})$ 
 such that the resulting state is $a_{\nu1}^{-} \ket{\bar{\psi}_1^{-}} - ia_{\nu2}^{-} \ket{\bar{\psi}_2^{+}}$. 
 By a measurement of the energy the probabilities are given by $|a_{\nu1}^{-}|^2$ for the outcome $E_{-}$ and $|a_{\nu2}^{-}|^2$ for $E_{+}$. 
A similar protocol holds for obtaining 
 $[\tilde{U}^{+}(\bar{U}^{+})^\dagger]_{\nu\mu}=a_{\nu\mu}^{+}$ 
 starting from
 $\ket{\tilde{\psi}_\nu^{+}}=a_{\nu 1}^{+} \ket{\bar{\psi}_1^{+}}+a_{\nu 2}^{+} \ket{\bar{\psi}_2^{+}}$.

However one has also to determine the phase of $a_{\nu\mu}^\pm$ to fully determine the basis transformation, 
for this one needs an internal rotation within the degenerate energy level. 
This can be achieved for example by the Wilzcek-Zee phase \cite{wilczek1984appearance}. 
Applying a Hadamard Gate (within the degenerate subspace) on the state $\ket{\tilde{\psi}_\nu^{-}}=a_{\nu 1}^{-} \ket{\bar{\psi}_1^{-}}+a_{\nu 2}^{-} \ket{\bar{\psi}_2^{-}}$ 
results in the state $(a_{\nu 1}^{-}+a_{\nu 2}^{-})/\sqrt{2} \ket{\bar{\psi}_1^{-}}+(a_{\nu 1}^{-}-a_{\nu 2}^{-})/\sqrt{2} \ket{\bar{\psi}_2^{-}}$. 
If we then perform again a two parameter modulation pulse as before with the duration $T=n(\pi/\Omega_1^{\pi/2})\approx (m+1/2)(\pi/\Omega_2^{\pi/2})$ 
the state ends up as 
$(a_{\nu 1}^{-}+a_{\nu 2}^{-})/\sqrt{2} \ket{\bar{\psi}_1^{-}}-i(a_{\nu 1}^{-}-a_{\nu 2}^{-})/\sqrt{2} \ket{\bar{\psi}_2^{+}}$. 
The probabilities of the energy measurement are then given by $1/2 + |a_{\nu 1}^{-}| |a_{\nu 2}^{-}| \cos(\varphi_{\nu1}-\varphi_{\nu2})$ 
for the outcome $E_{-}$ and $1/2- |a_{\nu 1}^{-}| |a_{\nu 2}^{-}| \cos(\varphi_{\nu1}-\varphi_{\nu2})$ for the outcome $E_{+}$ 
with $\varphi_{\nu\mu}$ being the phase of $a_{\mu\nu}^{-}=|a_{\mu\nu}^{-}|e^{i\varphi_{\mu\nu}}$. 
With this procedure one can ultimately determine all the coefficients $a_{\mu\nu}^{\pm}$ 
of the transformation by repeating this procedure for each state $\ket{\tilde{\psi}_\nu^\pm}$.

On the other hand for $\phi=0$ the terms of Eq.~(\ref{eq:7}) $e^{\pm i\phi}Q_{jk}^\pm+e^{\mp i\phi}Q_{kj}^\pm=2g_{jk}^\pm$ results in the quantum metric with the respective Rabi-frequencies
\begin{align}
\Omega^{0}_{\nu}=\frac{A}{\hbar}\left[\bar{Q}_{jj}^{\pm}+\bar{Q}_{kk}^{\pm}+2\bar{g}_{jk}^\pm\right]_{\nu\nu}^{1/2}\, .
\end{align}
Again the same condition as for the Berry curvature can be used to determine the quantum metric $g_{jk}$ with $\frac{A^2}{\hbar^2}\left[\bar{Q}_{jj}^{\pm}+\bar{Q}_{kk}^{\pm}+2\bar{g}_{jk}^\pm\right]_{\nu\mu}=\delta_{\nu\mu}(\Omega^{0}_{\nu})^2$.

\begin{figure}[t]
	\includegraphics[width=0.45\textwidth]{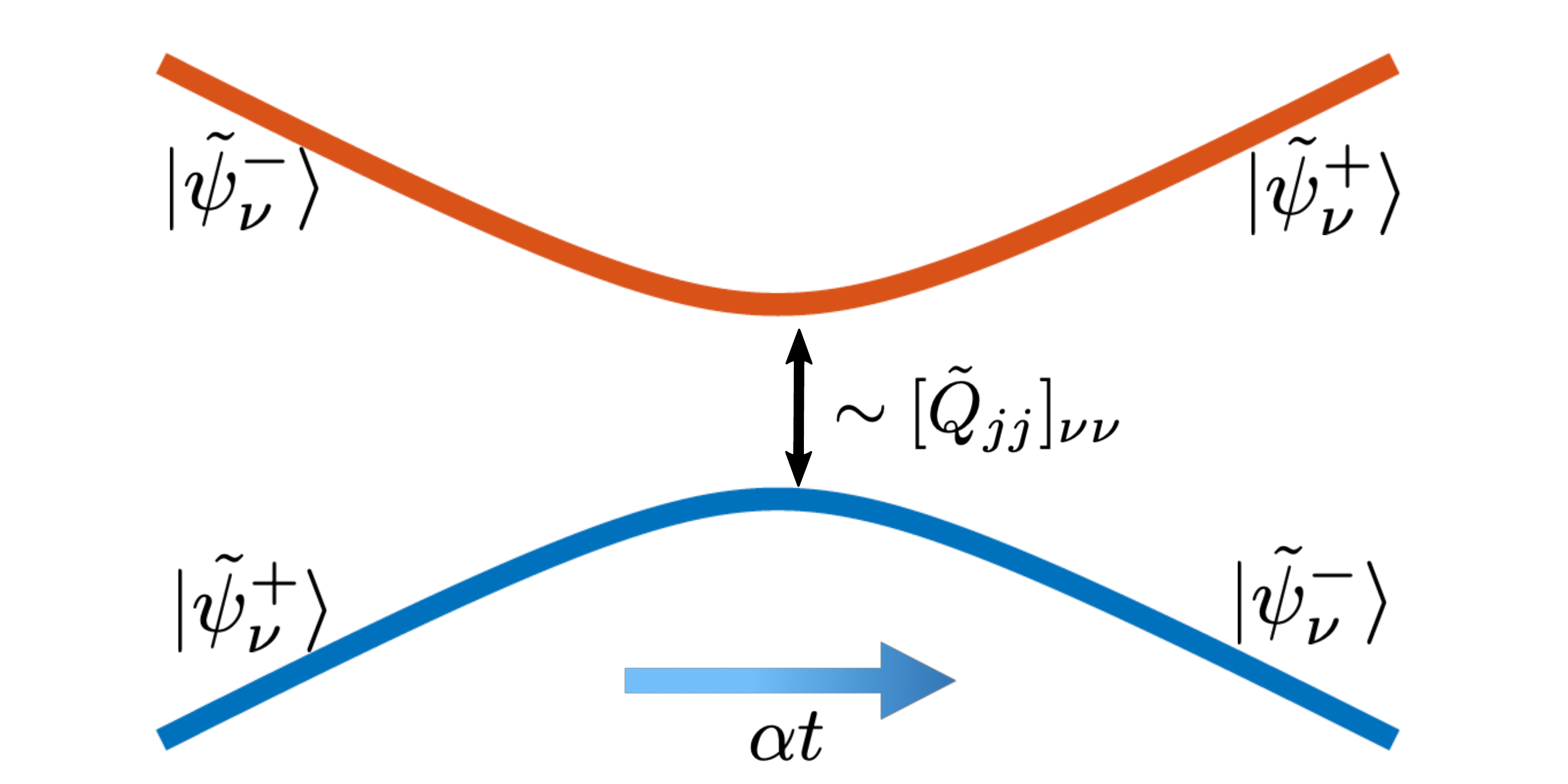}
	\caption{Schematic for a Landau-Zener transition for a pair $\ket{\tilde{\psi}_{\nu}^\pm}$ for a single modulation of the external parameter (or similar $\ket{\bar{\psi}_{\nu}^\pm}$ for a two parameter modulation). Starting from a state far away from the avoided crossing the effective two state system is linearly driven through this crossing proportional to the QGT $[\tilde{Q}_{jj}]_{\nu\nu}$. The final probability $P_{LZ}$ of remaining in the initial state is thus given by the eigenvalue of the QGT $[\tilde{Q}_{jj}]_{\nu\nu}$ (or $[\bar{Q}_{jj}^\pm+\bar{Q}_{kk}^\pm+e^{\pm i\phi}\bar{Q}_{jk}^\pm+e^{\mp i\phi}\bar{Q}_{kj}^\pm]_{\nu\nu}$ for the two parameter modulation).}\label{Fig3}
\end{figure}

\textit{Geometrical Landau-Zener transitions.}--
Another approach to determine the geometric properties is driving the system through an avoided crossing, see Fig.~\ref{Fig3}, 
while maintaining the geometrical Rabi oscillations as discussed above.
Essentially this can be described by an additional time dependent Zeeman like field, which is added to the RWA Hamiltonian such that
\begin{align}
H_{i}^{LZ}= H_{i}^{RWA} 
-
\alpha t
\sum_{\nu=1}^{N} \sum_{\sigma=\pm} \sigma \ket{\psi_\nu^{\sigma}}\bra{\psi_\nu^{\sigma}}\, ,
\end{align}
with $i=1$ for the single modulation, $i=2$ for the two parameter modulation, and $\alpha$ describing the (linear) tuning of the Zeeman like field.
For the Landau-Zener transition a two state system starts far away from the avoided crossing at  $t=-\infty$ and is tuned through this point at $t=0$ to $t=\infty$ where the final state of the system is evaluated. 
For a two-state system starting in one state the Landau-Zener probability $P_{LZ}$ gives the probability to remain in the same state \cite{zener1932non,wittig2005landau}. 
In our case we have several interacting two-state systems as discussed in \cite{vasilev2007degenerate}, which we drive through the avoided crossings. 
However if the system is prepared in an eigenstate of the quantum geometric tensor the system acts effectively as a two state system interacting between the pair $\ket{\tilde{\psi}_\nu^\pm}$ as depicted schematically in Fig.~\ref{Fig3}.
The interaction to all other states $\ket{\tilde{\psi}_\mu^\pm}$ with $\mu\neq \nu$ remains zero during the whole transition through the avoided crossing. 
Thus the result of the Landau-Zener transition can be directly applied to this effective two state system with the Landau-Zener probability directly proportional to the geometric perturbation $1-P_{LZ}^{\nu}=1-e^{-\pi |V_{\nu}^{i}|^2/\alpha}\approx \frac{\pi|V_{\nu}^{i}|^2}{\alpha}$ for $\alpha\gg |V_{\nu}^{i}|^2$ and with $|V_{\nu}^{1}|^2=\frac{A^2}{\hbar^2}[\tilde{Q}_{jj}]_{\nu\nu}$ for the single parameter drive and $|V_{\nu}^{2}|^2=\frac{A^2}{\hbar^2}[\bar{Q}_{jj}^\pm+\bar{Q}_{kk}^\pm+e^{\pm i\phi}\bar{Q}_{jk}^\pm+e^{\mp i\phi}\bar{Q}_{kj}^\pm]_{\nu\nu}$ for the two parameter drive.

\textit{Discussion.}-- 
We presented a new method to extract the quantum geometric tensor in non-Abelian systems with the help of geometric Rabi oscillations. 
In general this is not limited to specific systems, e.g. in electronic systems a modulation of the electric field can be applied to modulate the momentum of the electrons to extract geometric properties. 
Another possibility is to use this method in Josephson matter systems \cite{weisbrich2021second,riwar2016multi,eriksson2017topological,xie2017topological,meyer2017nontrivial,xie2018weyl,deb2018josephson,xie2019topological,klees2020microwave,klees2021ground}, 
where the superconducting phases play the role of external parameters defining the geometry of Andreev Bound states. 
In these systems the superconducting phases can be controlled by tuning magnetic fluxes, such that the presented method can be readily applied.

In a similar fashion this can be also applied in topological Josephson circuits \cite{peyruchat2021transconductance,fatemi2021weyl}, where the fluxes in the circuits can be modulated.

In the vicinity of Weyl points, where the bands are nearly degenerate, the RWA breaks down and no Rabi-oscillations occur. 
This issue underlines the critical nature in geometry of Weyl points in case of the presented work of the geometrical Rabi oscillations.

In general, systems are not limited to two bands, as presented in the work here. 
Assuming different energy spacing in multi band systems, the method is still applicable, as the driving frequency selects the respective band transition. 
This utility from the already well established method of Rabi oscillations is thus a great tool for exploring quantum geometry and topological properties in non-Abelian systems.

The authors acknowledge useful discussions with Guido Burkard and funding provided by the Deutsche Forschungsgemeinschaft (DFG, German Research Foundation) Grant No. RA 2810/1 and SFB 1432 – Project-ID 425217212.\\

\end{document}